\documentclass[aps,prd,twocolumn,floatfix,nofootinbib,showpacs,superscriptaddress]{revtex4-1}

\usepackage{amsmath,amsfonts,amssymb,bm}
\usepackage{graphicx}
\usepackage{subfigure}
\usepackage{color}
\definecolor{purple}{rgb}{0.5,0,0.5}
\definecolor{blue}{rgb}{0.0,0,0.9}
\usepackage[colorlinks=true, pdfstartview=FitV, linkcolor=purple, citecolor= purple, urlcolor=blue]{hyperref}

\begin{document}
\title{QCD phase transitions via a refined truncation of Dyson-Schwinger equations}

\author{Fei Gao }
\affiliation{Department of Physics and State
Key Laboratory of Nuclear Physics and Technology, Peking
University, Beijing 100871, China}

\author{Yu-xin Liu}
\email[Corresponding author: ]{yxliu@pku.edu.cn}
\affiliation{Department of Physics and State Key Laboratory of
Nuclear Physics and Technology, Peking University, Beijing 100871,
China}
\affiliation{Collaborative Innovation Center of Quantum Matter, Beijing 100871, China}
\affiliation{Center for High Energy Physics, Peking
University, Beijing 100871, China}

\date{\today}

\begin{abstract}
We investigate both the chiral and deconfinement phase transitions of QCD matter in a refined scheme of Dyson-Schwinger equations, which have been shown to be successful in giving the meson mass spectrum and matching the interaction with the results from {\it ab initio} computation.
%
%
%
We verify the equivalence of the chiral susceptibility criterion with different definitions for the susceptibility and confirm that the chiral susceptibility criterion is efficient to fix not only the chiral phase boundary but also the critical end point (CEP), especially when one could not have the effective thermodynamical potential. We propose a generalized Schwinger function criterion for the confinement. We give the phase diagram of both phase transitions and show that in the refined scheme  the position of the CEP shifts to lower chemical potential and higher temperature. Based on our calculation and previous results of the chemical freeze out conditions, we propose that the CEP locates in the states of the matter generated by the Au--Au collisions with $\sqrt{s_{NN}^{}}=9\sim15$ GeV.
\end{abstract}

\pacs{25.75.Nq, 11.10.Wx, 12.38.Lg, 21.65.Qr }

\maketitle

\section{Introduction}

The phase diagram of strong interaction matter in the plane of temperature $T$ and chemical potential $\mu$ has been investigated for a long time (for recent reviews, see {\it e.g.},  Refs.~\cite{BraunMunzinger:2009RMP,Fukushima:2011RPP,Owe:2013PPNP}).
The strong interaction is described by QCD which includes two important features: dynamical chiral symmetry breaking  and confinement, thusly the phase transitions are denoted as the QCD phase transitions and classified into two kinds: the ``dynamical chiral symmetry breaking (DCSB)--chiral symmetry (CS)" phase transition and the ``confinement(C)--deconfinement (DC)" phase transition.
To determine the order and the phase boundary of the chiral phase transition, thousand works have taken the chiral susceptibility criterion to carry out the investigations (see, {\it e.g.}, Refs.~\cite{Pisarski:1984PRD,Karsch:1994PRD,Hatsuda:1994PR,Fodor:2002PLB,Fodor:20024JHEP,DElia:2003PRD,Bernard:2005PRD,Aoki:2006Nature,Aoki:2006PLB,Allton:2006PRD,Forcrand:20023NPB,Ghosh:2006PRD,Hatta:2003PRL,Schaefer:2007PRD,Zhao:2008EPJC,Fodor:2009ax,Endrodi:2011JHEP,Kaczmarek:2011PRD,Karsch:2011PLB,Cea:2014PRD,Ding:2014PRL,Gavai:2005PRD,Schmidt:2008JPG,Li:2011PRD,Bazavov:2012PRD,Gupta:2014PRD,deForcrand:2014PRL,Fischer:2009PRD,Qin:2011PRL,Fischer:2013PLB,Bashir:2014JPG,Fischer:2014PRD,Fischer:2014NPA,Zong:2014FBS,Gao:2016PRD,Sasaki:2007PRD,Sasaki:2008PRD,Costa:2008PRD,Fu:2008PRD,Ciminale:2008PRD,Fukushima:2008PRD,Abuki:2008PRD,Schaefer:2009PRD,Jiang:2013PRD,Xin:2014PRDa,Mohanty:2009NPAc,Lacey:2014PRL,Lacey:2015PRL}). %
Since the system involves usually multiple variables, in turn, there are different definitions for the chiral susceptibility. The equivalence between the differently defined susceptibilities in signalling the phase transition has not yet been examined thoroughly.
Even the equivalence of the susceptibility criterion to the thermodynamical potential criterion has not yet been clarified either, except in some simple models (see, {\it e.g.}, Ref.~\cite{Zhao:2008EPJC}).
For the confinement--deconfinement phase transition, since the confinement is defined as that the color degrees of freedom are confined inside hadrons and could not be observed as isolated states.
It can naturally be represented by the violation of the positivity of the spectral density function.
The positivity of the spectral density function is then sufficient to label the deconfinement~\cite{Roberts:1994PPNP,Alkofer:2001PR}.
Because of the difficulties in calculating the spectral function, one usually evaluates the Schwinger function which is defined as the Fourier transformation of the propagator (at finite temperature and finite chemical potential)~\cite{Bender:1996PRL,Bender:1998PLB,Mueller:2010EPJC,Qin:2011PRD,Gao:2014PRD,Qin:2013PRD,Bashir:2008PRC,Bashir:2009FBS}. Nevertheless the Schwinger function criterion  fails in some cases~\cite{Alkofer:2004PRD},
because the Schwinger function is the integral of the spectral density.
It is then necessary to extend the Schwinger function so that the criterion is equivalent to that of the spectral density function, and the numerical calculation  is easy to carry out.

It has been well known that the QCD phase transitions happen at the energy scale $10^{2}\;$MeV,
one must take nonperturbative QCD approaches to accomplish the investigations.
Therefore lattice QCD simulations have been widely implemented (see, {\it e.g.}, Refs.~\cite{Owe:2013PPNP,Karsch:1994PRD,Fodor:2002PLB,Fodor:20024JHEP,DElia:2003PRD,Bernard:2005PRD,Aoki:2006Nature,Aoki:2006PLB,Allton:2006PRD,Forcrand:20023NPB,deForcrand:2014PRL,Fodor:2009ax,Endrodi:2011JHEP,Kaczmarek:2011PRD,Cea:2014PRD,Gavai:2005PRD,Schmidt:2008JPG,Li:2011PRD,Bazavov:2012PRD,Gupta:2014PRD,Karsch:2011PLB,Ding:2014PRL}). However, the ``sign problem"~\cite{Owe:2013PPNP}  retards it making great progress in large chemical potential region.
The Dyson-Schwinger (DS) equation  method~\cite{Roberts:1994PPNP,Alkofer:2001PR,Roberts:2000PPNP,Maris:2003IJMPE,Fischer:2006JPG,Bashir:2012CTP,Cloet:2013PPNP} and the functional renormalization group approach~\cite{Pawlowski:2007AP}, which include both the DCSB and the confinement inherently~\cite{McLerran:2007NPA,Pawlowski:2013PRD}, have thus played the role.
Not only the general features of the phase transitions but also the CEP, the property of the state in the temperature region above but near the pseudo-critical one, even the baryon number fluctuations and some transport properties have then been obtained (see, {\it e.g.},  Refs.~\cite{Fischer:2009PRD,Qin:2011PRL,Fischer:2013PLB,Zong:2014FBS,Bashir:2014JPG,Fischer:2014PRD,Fischer:2014NPA,Bender:1996PRL,Bender:1998PLB,Mueller:2010EPJC,Qin:2011PRD,Gao:2014PRD,Qin:2013PRD,Bashir:2008PRC,Bashir:2009FBS,Blaschke:1998PLB,Maris:2003EPJA,Chen:2008PRD,Fischer:2009PRLa,Xin:2014PRDb,Eichmann:2016PRD,Gao:2016PRD,Chen:2015mga,Pawlowski:2013PRD,Pawlowski:2015PRDa,Pawlowski:2015PRDb,Pawlowski:2015PRL,Pawlowski:20156PRD}). The DS equation approach of QCD is  a method of continuum quantum field theory.
It is  convenient to stretch the calculations on the whole $\mu$--$T$ plane  without further approximation. This advantage makes it better to obtain the information of the phase structure on $\mu$--$T$ plane than the lattice QCD simulations at present stage, especially in case of large chemical potential.
However, almost all the previous work via the DS equation approach were based on the bare approximation for the quark--gluon interaction vertex.
On hadron property side, which is usually taken as the calibration to fix the parameter(s) in the DS equation approach, it has been shown that the bare vertex truncation leads to it only accurate for ground-state vector- and isospin-nonzero-pseudoscalar mesons~\cite{Maris:2003IJMPE,Bashir:2012CTP,Cloet:2013PPNP}  because corrections in these channels largely cancel each other owing to the parameter-free preservation of the Ward-Green-Takahashi (WGT) identities \cite{Ward:1950PR,Green:1953PPSA,Takahashi:1957NC}.
The corrections do not cancel in other channels~\cite{Roberts:1996jx,Roberts:1997PPN,Bender:2002PRC,Bhagwat:2004PRC},
studies based on such a truncation have thus provided usually poor results for scalar, axial-vector and exotic state mesons~\cite{Burden:1997PRC,Watson:2004FBS,Fischer:2009PRLb,Krassnigg:2009PRD,Qin:2011PRC,Qin:2012PRC}, %
and exhibited gross sensitivity to model parameters for excited states~\cite{Qin:2011PRC,Qin:2012PRC,Holl:2004PRC,Holl:2004IJMPA}
and tensor mesons~\cite{Krassnigg:2011PRD}.

A recently developed truncation scheme~\cite{Chang:2011PRL,Chang:2012PRC} has been found to have bridged the bottom--up scheme with the {\it ab initio} computation in continuum QCD~\cite{Binosi:2015PLB}.
The scheme preserves the WGT identities~\cite{Qin:2013PLB}, and introduces the DCSB effect into the interaction kernel~\cite{Chang:2011PRL}.
It has successfully given realistic hadron properties for not only the ground states of axial-vector mesons but also some exited state mesons~\cite{Chang:2012PRC}.
Considering the success in describing the hadron properties of the new scheme and the coincidence with the {\it ab initio} computation, it is imperative to implement the refined scheme to reanalyze the phase transitions in the $T$--$\mu$ plane and examine the discrepancy of the results induced by the difference of the  truncation schemes.

We take then the refined truncation scheme to investigate the QCD phase transitions in this paper.
After analyzing the chiral susceptibility criteria for the chiral phase transition and generalizing the  Schwinger function criterion for the confinement phase transition,
we obtain the phase diagrams of the transitions.
We show that the two kind phase transitions coincide with each other and the results of the chiral susceptibility criteria with different definitions for the susceptibility deviate only in the crossover region slightly due to the nature of the crossover.
We  verify that the phase transition temperature at zero chemical potential is (at least) $150\;$MeV which is consistent with the lattice QCD results, and propose that the critical end point (CEP) of the chiral phase transition locates  at $(\mu_{B}^{},T)=(262,\, 126)\,$MeV. It indicates that the new truncation decreases the chemical potential of the CEP. Such a location of the CEP is in the range of the states being able to generated with the $\sqrt{s} \cong 9 \sim 15\;$GeV Au--Au collision with the parametrization of the energy dependence of $\mu$~\cite{Cleymans:2006PRC,Chen:2015mga,Andronic:2006NPA,Borsanyi:2014PRL}, in turn, may be observed in the beam energy scan
experiments at RHIC~\cite{Adamczyk:2014PRL,Luo:2015ewa}.

The remainder of this paper is organized as follows.
In Sec. \uppercase\expandafter{\romannumeral2} we reiterate briefly the DS equation approach and its refined truncation scheme.
In Sec. \uppercase\expandafter{\romannumeral3} we analyze the criteria of the chiral phase transition,
show their equivalence and emphasize the efficiency of the chiral susceptibility criterion in fixing the CEP.
In Sec. \uppercase\expandafter{\romannumeral4}, we extend the Schwinger function criterion
and show the equivalence between the generalized Schwinger function and the spectral density function.
In Sec. \uppercase\expandafter{\romannumeral5} we give our results of the phase diagrams
and discuss the properties.
Finally, we summarize in Sec.\uppercase\expandafter{\romannumeral6}.

\section{Quark Gap Equation }
\label{DSEs}

In the DS equation approach of QCD, the quark propagator $S$ at finite temperature and quark chemical potential can be determined with the gap equation
\begin{eqnarray}
\label{eq:gap1}
S(\vec{p},\tilde\omega_n)^{-1} &=&  i\vec{\gamma}\cdot\vec{p}
+ i\gamma_4 \tilde \omega_{n} + m_{0}^{}
  + \Sigma(\vec{p}, \tilde\omega_{n}) \, ,  \\
\nonumber
\Sigma(\vec{p},\tilde\omega_n) &=& T\sum_{l=-\infty}^\infty \! \int\frac{d^3{q}}{(2\pi)^3}\; {g^{2}} D_{\mu\nu} (\vec{p}-\vec{q}, \Omega_{nl}; T, \mu) \quad  \\
& & \times \frac{\lambda^a}{2} {\gamma_{\mu}} S(\vec{q},
\tilde\omega_{l}) \frac{\lambda^a}{2}
\Gamma_{\nu} (\vec{q}, \tilde\omega_{l},\vec{p},\tilde\omega_{n})\, ,
\label{eq:gap2}
\end{eqnarray}
where $m_{0}^{}$ is the current quark mass, $\tilde\omega_{n}^{} = \omega_{n}^{} + i \mu$ with $\omega_n=(2n+1)\pi T$ being the quark Matsubara frequency, and $\mu$ the quark chemical potential, $\Omega_{nl} = \omega_{n} - \omega_{l}$;
$ {g^{2}} D_{\mu\nu} (\vec{p}-\vec{q}, \Omega_{nl}; T, \mu) $ is the interaction with $D_{\mu \nu }^{}$  the dressed-gluon propagator; and $\Gamma_{\nu}$  the dressed-quark-gluon vertex.

The gap equation's solution can be decomposed as
\begin{eqnarray}\label{eq:qdirac}
\nonumber
S(\vec{p},\tilde\omega_n)^{-1} & = & i\vec{\gamma} \cdot \vec{p}\, A(\vec{p}\,^2, \tilde\omega_{n}^2) \\
&& + i\gamma_{4} \tilde\omega_{n} C(\vec{p}\,^2, \tilde\omega_{n}^2) + B(\vec{p}\,^2, \tilde\omega_{n}^2) \, .   \quad
\end{eqnarray}
The interaction has generally the form
\begin{equation}
g^2 D_{\mu\nu}(\vec{k}, \Omega_{nl}) = P_{\mu\nu}^{T} D_{T}(\vec{k}\,^2, \Omega_{nl}^2) + P_{\mu\nu}^{L} D_{L}(\vec{k}\,^2, \Omega_{nl}^2)\,,
\end{equation}
where $P_{\mu\nu}^{T,L}$ are, respectively, the transverse and longitudinal projection operators.
As temperature changes, the transverse and longitudinal part is found  different in the practical computation~\cite{Silva:2014PRD}, however, it's still a good approximation to set $D_{T}^{}=D_{L}^{}$
to study the phase structure of QCD~\cite{Qin:2011PRL}.
We take then the approximation $D_{T}^{}=D_{L}^{}  = {\mathcal{D}}$ in this paper.
Modern DS equation and lattice QCD studies indicate that the gluon propagator is a bounded, regular function of spacelike momenta, which achieves its maximum value on the domain at $k^2=0$~\cite{Bowman:2004PRD,Bogolubsky:2009PLB,Boucaud:2010PRD,Oliveira:2011JPG,Cucchieri:2012PRD,Aguilar:2012PRD,Ayala:2012PRD,Dudal:2012PRD,%
Strauss:2012PRL,Zwanziger:2013PRD,Blossier:2013PRD}. We then employ an interaction which  expresses these features~\cite{Qin:2011PRC,Chang:2012PRC} as :
\begin{eqnarray}
\label{gluon}
\mathcal{D}({k^{2}_{\Omega}}, {m_{g}^{2}}) & = & 8{\pi^{2}} {\mathcal{D}}
\frac{1}{\omega^{4}} e^{-{s_{\Omega}^{}}/\omega^{2}} \notag\\
& & + \frac{8{\pi^{2}} {\gamma_{m}}}{{\ln}[ \tau \! + \! (1 \! + \!
{s_{\Omega}^{}}/{\Lambda_{\text{QCD}}^{2}} ) ^{2} ] } \,
{\cal F}(s_{\Omega}^{}) \, ,
\end{eqnarray}
with  $({\mathcal{D}}\omega)^{1/3}=0.52$ GeV and $\omega=0.5$ GeV, ${\cal F}(s_{\Omega}) = (1-\exp(-s_{\Omega}/4 m_{t}^{2}))/s_{\Omega}$,  $\tau=e^2-1$, $m_t=0.5\,$~GeV, $\gamma_m=12/25$, $\Lambda^{}_{\text{QCD}}=0.234\;$GeV, and $s_{\Omega}^{} = \Omega^2 + \vec{k}^{2} $.

We then implement  the refined truncation to the quark-gluon interaction vertex~\cite{Chang:2011PRL}  which would be called the ACM kernel or CLR kernel, which reads:
\begin{eqnarray}
 \label{eq:ACMvertex}
\Gamma_{\mu} &=& \Gamma_{\mu}^{\textrm{BC}}+\Gamma_{\mu}^{\textrm{ACM}} \, , \\
\Gamma_{\mu}^{\textrm{ACM}}&=&\Gamma_{\mu}^{\textrm{ACM}_{4}}
+\Gamma_{\mu}^{\textrm{ACM}_{5}} \, .
 \end{eqnarray}

The longitudinal part of this vertex is just the Ball--Chiu (BC)  vertex~\cite{Ball:1980PRD}, which it is the unique solution constrained by the transverse WGT identities~\cite{Qin:2013PLB}.
The generalized BC vertex at finite temperature\cite{Qin:2011PRL,Maris:2001PRC,Chen:2008PRD} reads:
\begin{eqnarray}
 \label{eq:BCvertex}
\Gamma_{\mu}^{\textrm{BC}}(\vec{q},\tilde\omega_{l}^{},\vec{p},\tilde\omega_{n}^{}) &=& \gamma^{T}_{\mu}\Sigma_{A}^{} +\gamma^{L}_{\mu}\Sigma_{C}^{} \notag \\
 &&+(p_{n}^{} + q_{l}^{})_{\mu} \Big[ \frac{1}{2}\gamma^{T}_{\alpha}
 (p_{n}^{} + q_{l}^{})_{\alpha}^{} \Delta_{A}^{}  \quad \notag \\
 &&+\frac{1}{2}\gamma^{L}_{\alpha} (p_{n}^{}+q_{l}^{})_{\alpha}^{} \Delta_{C}^{} - i \Delta_{B}^{} \Big] \, ,
 \end{eqnarray}
 with
\begin{eqnarray}
& & p_{n}^{} =(\vec{p},\tilde\omega_{n}^{}), \qquad  q_{l}^{} =(\vec{q},\tilde\omega_{l}^{}) ,  \notag \\
& & \Sigma_{F}(\vec{q}^{2}, \tilde\omega^{2}_{l}, \vec{p}^{2}, \tilde\omega^{2}_{n}) =\frac{1}{2}[F(\vec{q}^2,\tilde\omega^{2}_{l})+F(\vec{p}^{2},\tilde\omega^{2}_{n})]\,, \notag\\
& & \Delta_{F}(\vec{q}^{2},\tilde\omega^{2}_{l},\vec{p}^{2},\tilde\omega^{2}_{n})  =  \frac{F(\vec{q}^{2},\tilde\omega^{2}_{l}) - F(\vec{p}^{2},\tilde\omega^{2}_{n})}{q^{2}_{l} - p^{2}_{n} } \, ,
 \end{eqnarray}
 where $F= A, B, C$, and $\gamma^{T}_{\mu} = \gamma_{\mu}^{} - \gamma_{\mu}^{L}$,  $\gamma^{L}_{\mu} = u_{\mu}^{} \gamma_{\alpha}^{} u_{\alpha}^{}$ with $u=(0,0,0,1)$.

The part $\Gamma^{\textrm{ACM}}_{\mu}$ is the transverse structure in the vertex that characterizes the DCSB effect in the quark-gluon vertex through the anomalous chromomagnetic moments (ACM)~\cite{Chang:2011PRL}, which reads:
\begin{eqnarray}
 \label{eq:ACMvertex}
\Gamma_{\mu}^{\textrm{ACM}_{4}^{}}&=&[T_{\mu\nu} l_{\nu}^{} \gamma \cdot k+i  T_{\mu\nu}\gamma_{\nu}^{} \sigma_{\rho\sigma}^{} l_{\rho}^{} k_{\sigma}^{} ]
\tau_{4}^{}(p_{n}^{} , q_{l}^{}) \, , \quad \\
\Gamma_{\mu}^{\textrm{ACM}_{5}^{}}&=&\sigma_{\mu\nu}^{} k_{\nu}^{}
\tau_{5}^{} (p_{n}^{} , q_{l}^{} ) \, , \\
\tau_{4}^{} &=& \frac{2 \tau_{5}^{}(p_{n}^{}, q_{l}^{}) [2(M(p_{n}^{2})+M(q_{l}^{2}))]}{p_{n}^{2}+M(p_{n}^{2})^{2} + q_{l}^{2}+M(q_{l}^{2})^{2}} \, ,
 \end{eqnarray}
where $\sigma_{\mu \nu}^{} = (i/2)[\gamma_{\mu}^{} , \gamma_{\nu}^{} ]$,
$T_{\mu\nu}^{} =\delta_{\mu\nu}^{} - k_{\mu}^{} k_{\nu}^{}/k^{2}$, $k_{\nu}^{}=(p_{n}^{} - q_{l}^{})_{\mu}^{}$, $l_{\mu}^{} = \frac{(p_{n}^{} + q_{l}^{})_{\mu}^{}}{2}$,  \,
$\tau_{5}^{} =\eta\Delta_{B}^{} $ with $\eta$ a parameter, and  $M(x)=B(x)/A(x)$.

This  truncation  scheme satisfies both the longitudinal and transverse WGT identities~\cite{Qin:2013PLB}, and  makes the interaction match with the results obtained from {\it ab initio} computation~\cite{Binosi:2015PLB}.

\section{Chiral Phase Transition}

\subsection{Criteria for chiral phase transition}

To investigate the chiral phase transition, the chiral susceptibilities have commonly been taken as the criteria. The (generalized) chiral susceptibility is defined  as the derivative of the chiral order parameter with respective to the control parameters, such as the current quark mass $m_{0}^{}$, temperature $T$ or chemical potential $\mu$.
The susceptibility can be connected with the thermodynamical potential.
To this end, we consider the Cornwall--Jackiw--Tomboulis (CJT) effective thermodynamical potential for quarks, which reads~\cite{Cornwall:1974PRD}:
\begin{equation}   \label{eq:CJTP}
\Gamma(S)=-Tr[{\ln}(S^{-1}_{0}S)-S^{-1}_{0}S+1]+\Gamma_{2}(S),
\end{equation}
where $S_0$ stands for the bare quark propagator, $\Gamma_{2}$ is the 2PI contribution. Calculating the variation with respective to quark propagator, we have:
\begin{eqnarray}
&&\frac{\partial \Gamma}{\partial S}=-S^{-1}+S^{-1}_{0}+\frac{\partial \Gamma_{2}(S)}{\partial S}\, , \label{eq:CJT}\\
&&\frac{\partial^2 \Gamma}{\partial S^2} = S^{-2} + \frac{\partial^{2} \Gamma_{2}(S)}{\partial S^{2}} \, .
\end{eqnarray}

The quark propagator's DS equation could be derived through the extreme condition of Eq.~(\ref{eq:CJT}). Meanwhile, if calculating the  derivative of the extreme condition with respective to the current quark mass for quark propagator's  DS equation, we obtain
\begin{equation}
-S^{-2}\frac{\partial S}{\partial m_{0}^{}}=1+\frac{\partial^{2} \Gamma_{2}(S)}{\partial S^{2}}\frac{\partial S}{\partial m_{0}^{}},\label{eq:suscep}
\end{equation}

The function $S$ represents the dynamical symmetry breaking, and can be considered as the order parameter. The $\frac{\partial S}{\partial m_{0}^{}} $ can then be regarded as the generalized chiral susceptibility, too.
Comparing Eq.~(\ref{eq:CJT}) and Eq.~(\ref{eq:suscep}) we can find straightforwardly the relation between the generalized chiral susceptibility and the thermodynamical potential
\begin{equation}
\label{eq:suspo}
\frac{\partial S}{\partial m_{0}^{}}=-\frac{1}{\partial^2 \Gamma/\partial S^2}.
\end{equation}
Similar relations between $\frac{\partial S}{\partial T}$, $\frac{\partial S}{\partial \mu}$ and thermodynamical potential can also be easily derived as above.

It is well known that  the sign changing of the second order derivative of the thermodynamical
potential at the state satisfied the extreme condition is conventionally regarded as the signature of
a phase transition.
When we consider the case in nonperturbative point of view completely, we could not have the thermodynamical potential explicitly. The conventional thermodynamical criterion fails unfortunately.
We should then develop new criteria.
From the above relations one can notice that the (generalized) chiral susceptibility is the reciprocal of the second order derivative of the thermodynamical potential (Similar relation in case of NJL model has been given in Ref.~\cite{Zhao:2008EPJC}). It means that the (generalized) chiral susceptibilities play the same role as the thermodynamical potential in identifying the chiral phase transition,  and can thus be taken as criteria for the transition.

It is also well known that, in case of chiral limit, the trace of the quark propagator is the chiral quark condensate $\langle \bar{q} q \rangle$. Extending such a definition to the case beyond chiral limit we have
\begin{eqnarray}
\langle \bar{q}q \rangle_{m_{0}^{}}^{} & = & Tr[S(p)_{m_{0}^{} \neq 0 }^{} ] \nonumber \\
& = & - Z_4 N_{c} N_{f} \int
\frac{d^{4}p}{(2\pi)^4} tr[S(p)_{m_{0}^{} \neq 0 }^{} ] \, ,    \label{eq:CondBcL}
\end{eqnarray}
where $Z_{4}$ is the renormalization constant, $N_{c}$ the number of colors and $N_{f}$ the number of flavors. However the direct trace of the quark propagator in case of nonvanishing current quark mass    contains quadratic divergence. Some subtraction schemes must be taken when calculating the condensate. Noticing that the quark mass function could be written as~\cite{Roberts:1994PPNP}:
\begin{eqnarray}
&&M(-Q^2)=\frac{B(-Q^2)}{A(-Q^2)}\xrightarrow{Q^2\rightarrow \infty}   \quad \notag  \\ &&\quad \frac{c}{Q^2} \Big{[} \ln{\! \big{(} \frac{Q^2}{\Lambda^{2}_{QCD}} \big{)} } \Big{]}^{\gamma_{m}^{} -1}
+{m_{0}^{}} \Big{[} \frac{\ln(\mu^2/\Lambda^2_{QCD})}{\ln(Q^2/\Lambda^2_{QCD})} \Big{]}^{\gamma_{m}^{} } , \quad
\end{eqnarray}
with  $c=-\frac{4\pi^{2} \gamma_{m}^{}}{3}\frac{ \langle \bar{q}q \rangle } {[\ln(\mu^{2}/\Lambda^2_{QCD})]^{\gamma_{m}^{}}}$ and $\mu$ is the renormalization scale,
one can find that the quadratic divergence is linearly dependent on the current quark mass $m_{0}^{}$. It is apparent that the quadratic divergence can be removed when the quark condensate is redefined as
\begin{equation}
\label{eq:conden}
\langle \bar{q}q \rangle = \langle \bar{q}q \rangle_{m_{0}^{}}^{} - {m_{0}^{}} \frac{\partial \langle \bar{q}q \rangle_{m_{0}^{}}^{}}{\partial m_{0}^{} \; } \, .
\end{equation}
It is evident that the quark condensate is a direct measure of the dynamical quark mass generation. This condensate has then commonly been taken as the chiral order parameter.
As the integral is extended to that at finite temperature and/or finite chemical potential, the responsibility $\frac{\partial \langle \bar{q} q \rangle}{\partial T}$,   $\frac{\partial \langle \bar{q} q \rangle}{\partial \mu}$,  $\frac{\partial \langle \bar{q} q \rangle}{\partial m_{0}^{}}$ and so on are also regarded as the chiral susceptibility and could naturally be taken as the signatures of the chiral phase transition,
which are much more concrete than the ones expressed similar as Eq.~(\ref{eq:suspo}).

To show more intuitively the equivalence of the chiral susceptibility criterion with the thermodynamical potential criterion, we recall the Eqs.~(\ref{eq:CJTP}), (\ref{eq:CondBcL}) and (\ref{eq:conden}).
It is apparent that the thermodynamical potential can be rewritten as
$$  \Gamma(S) = \Gamma(\langle \bar{q} q \rangle )  \, . $$
Along the line of Landau phase transition theory, the above thermodynamical potential of the state around the phase transition can be expanded in terms of the powers of the condensate as
\begin{equation}  \label{eq:effeTP-cond}
\Gamma (\langle \bar{q} q \rangle, \zeta ) = \Gamma_{0}^{}(\zeta) + \frac{1}{2} \alpha \langle \bar{q} q \rangle ^{2} + \frac{1}{4} \beta \langle \bar{q} q \rangle^{4} + \frac{1}{6} \gamma \langle \bar{q} q \rangle ^{6}
\, ,
\end{equation}
where $\zeta$ denotes all the controlling variables such as temperature $T$, chemical potential $\mu$ and so on, $\alpha$, $\beta$ and $\gamma$ are the interaction strength parameters. We have then
the other form of the generalized chiral susceptibility
\begin{equation}
\chi = \left( \frac{\partial \langle \bar{q} q \rangle}{\partial \zeta} \right)^{}_{\zeta = \zeta_{c}^{}}  \, ,
\end{equation}
where $\zeta_{c}^{}$ stands for the (pseudo-)critical condition of the phase transition.

After differentiating the stationary condition of the effective thermodynamical potential with respect to the controlling parameter, one can have easily
$$ \displaylines{\hspace*{2mm} \chi =
\frac{-\langle \bar{q} q \rangle \big( \frac{\partial \alpha}{\partial \zeta} \big)_{\zeta = \zeta_{c}^{}}
- \langle \bar{q} q \rangle^{3} \big( \frac{\partial \beta}{\partial \zeta} \big)_{\zeta = \zeta_{c}^{}}
- \langle \bar{q} q \rangle^{5} \big( \frac{\partial \gamma}{\partial \zeta} \big)_{\zeta = \zeta_{c}^{}}}
{\alpha + 3 \beta \langle \bar{q} q \rangle^{2} + 5 \gamma \langle \bar{q} q \rangle^{4} } \hfill{}
\cr \hspace*{5mm}
=
- \frac{\langle \bar{q} q \rangle \big( \frac{\partial \alpha}{\partial \zeta} \big)_{\zeta = \zeta_{c}^{} }
+ \langle \bar{q} q \rangle ^{3} \big( \frac{\partial \beta}{\partial \zeta} \big)_{\zeta = \zeta_{c}^{}}
+ \langle \bar{q} q \rangle^{5} \big( \frac{\partial \gamma}{\partial \zeta} \big)_{\zeta = \zeta_{c}^{}}}
{ ( \frac{\partial ^{2} \Gamma}{\partial \langle \bar{q} q \rangle^{2}})_{\frac{\partial \Gamma}{\partial \langle \bar{q} q \rangle } = 0} ^{} }  \, .
\hfill{} }
$$
It is also known that, for the symmetry restoration phase transition, the derivatives $(\frac{\partial \alpha}{\partial \zeta})_{\zeta = \zeta_{c}^{}}^{} > 0$, $(\frac{\partial \beta}{\partial \zeta})_{\zeta = \zeta_{c}^{}}^{} > 0$, $(\frac{\partial \gamma}{\partial \zeta})_{\zeta = \zeta_{c}^{}}^{} > 0$.
Since $\langle \bar{q} q \rangle < 0$, such a susceptibility $\chi$ takes the same sign as the $ ( \frac{\partial ^{2} \Gamma}{\partial \langle \bar{q} q \rangle^{2}})_{\frac{\partial \Gamma}{\partial \langle \bar{q} q \rangle } = 0} ^{} $.
One can then recognize that, if the chiral susceptibility is positive,
the state is in a stable phase, and the negative susceptibility stands for a unstable phase.
Therefore the susceptibilities  $ \frac{\partial \langle \bar{q} q \rangle }{\partial T}$
and $ \frac{\partial \langle \bar{q} q \rangle }{\partial \mu}$ have been commonly taken as the signatures of the chiral phase transition.

In practical calculation, since the dynamical mass of a quark is infrared-dominant,
the scalar part of the inverse quark propagator at zero momentum $B(0,\tilde\omega_0^2)$
is a good representation for  the chiral property of the quark.
We can see it clearly through introducing a cut-off $\Lambda\geq B(0)$ and simply estimating the mass function in quark propagator with $B(0)$. The integral of the quark propagator in calculating the quark condensate is then proportional to $B(0)\Lambda^2$.
Therefore the chiral susceptibility can be simply rewritten as~\cite{Qin:2011PRL}:
\begin{equation}
\label{eq:critiria}
\chi(0,\tilde\omega_0)=\frac{\partial}{\partial m_{0}^{}}B(0,\tilde\omega_0^2).
\end{equation}

In short, our above analysis indicates that the chiral susceptibility criterion is exactly equivalent to the thermodynamical potential criterion in analyzing a phase transition. Especially, in case of that one could not have the (effective) thermodynamical potential $\Gamma$ when considering completely the nonperturbative effect, the chiral susceptibility criterion can still work well.
In this paper, we consider two flavor quark system with degenerate bare mass $m_{0}^{} = 3.4\;$MeV to analyze the criteria of the chiral phase transition again and verify the equivalence of the different expressions of the chiral susceptibility in practical usage via calculations in the DS equation scheme with the refined quark--gluon interaction vertex.

\subsection{Numerical Results}

\begin{figure}[htb]
\centerline{\includegraphics[width=0.450\textwidth]{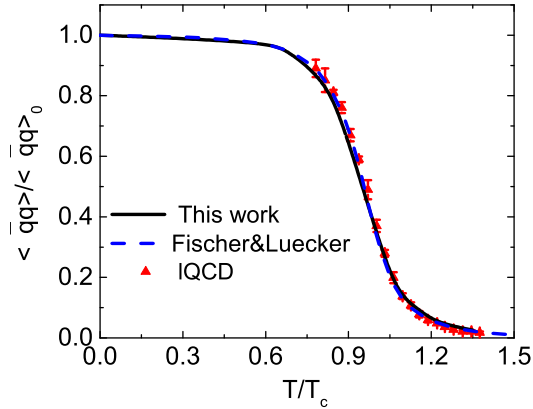}}
\caption{(color online) Calculated scaled quark condensate (solid line)  at $\mu=0$ as a function of $T/T_{c}(\mu=0)$ compared with the results from other DS equation calculations
(dashed line, taken from Ref.~\cite{Fischer:2014PRD}) and lattice QCD simulation (filled triangles with error bars, taken from Ref.~\cite{Borsanyi:2010JHEP}. ) }
\label{fig:conden1}
\end{figure}

\begin{figure}[htb]
\centerline{\includegraphics[width=0.450\textwidth]{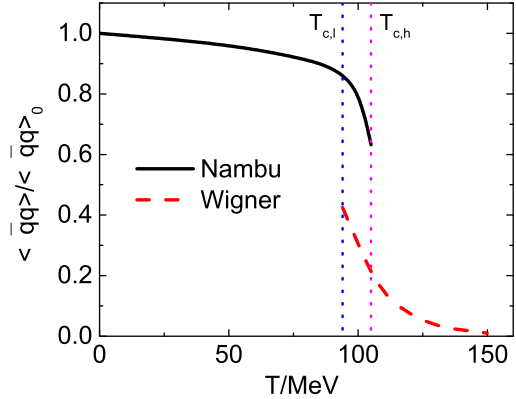}}
\centerline{\includegraphics[width=0.450\textwidth]{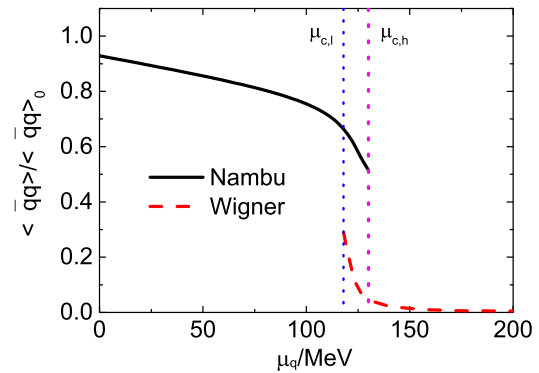}}
\caption{(color online)  Calculated scaled quark condensate at $\mu=150\;$MeV as a function of $T$ (\emph{upper panel}: \emph{Solid line}--Nambu phase; \emph{dashed line}--Wigner phase) and that at $T=110\;$MeV as a function of $\mu$ (\emph{lower  panel}: \emph{Solid line}--Nambu phase;\emph{dashed line}--Wigner phase).
} \label{fig:conden2}
\end{figure}

Firstly we calculate the quark condensate with Eq.~(\ref{eq:conden}).
The obtained result of the temperature dependence of the condensate at zero chemical potential scaled with that at zero temperature and the comparison with those given in other DS equation calculations~\cite{Fischer:2014PRD} and lattice QCD simulation~\cite{Borsanyi:2010JHEP} are shown in Fig.~\ref{fig:conden1}.
The obtained result of the temperature dependence of the scaled condensate at a sizeable quark chemical potential (150~MeV) and the quark chemical potential dependence of the scaled condensate at a finite temperature (110~MeV)  are shown in Fig.~\ref{fig:conden2}.

Looking over Fig.~\ref{fig:conden1}, one can easily recognized that our presently calculated temperature dependence of the condensate agrees with the lattice QCD simulation result and the previous DS equation calculation result excellently. In general, the condensate at low temperature barely changes as the temperature increases till about $T\sim 0.8\, T_{c}^{}$ with $T_{c}^{}$, the inflection points of corresponding curves (in our calculation, $T_{c}^{}=150.8\;$MeV).
Straightforwardly, $T_{c}^{}$ is the temperature for the $\frac{\partial \langle \bar{q}q \rangle}{\partial T}$, which is one kind of  definition of chiral susceptibility,  to reach its maximum. In turn, it is commonly regarded as the psuedo-critical temperature of the chiral phase transition (a crossover in case of beyond chiral limit).
In contrast to the continuous evolution at zero chemical potential, the Fig.~\ref{fig:conden2} manifests evidently that, in high chemical potential region, both the temperature and the chemical potential dependences of the quark condensate become discontinuous.
In more detail, the condensates for the Nambu solution and Wigner solution are separated in a special region, they ``jump to" each other at two distinct chemical potentials $\mu_{c,l}^{}$ and $\mu_{c,h}^{}$. These features indicate apparently that the phase transition becomes first order, and the region $\mu \in [\mu_{c,l} , \mu_{c,h}^{} ] $ and the counterpart of the temperature are just the coexistence regions.

\begin{figure}[htb]
\centerline{\includegraphics[width=0.450\textwidth]{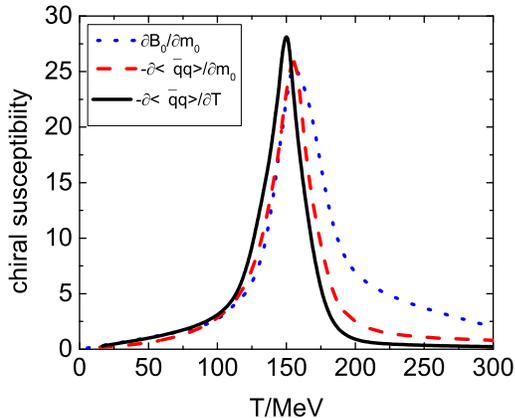}}
\caption{(color online) Calculated $\frac{\partial  \langle \bar{q}q \rangle}{\partial T}$,  $\frac{\partial  \langle \bar{q}q \rangle }{\partial m_{0}^{}}$ and $\frac{\partial B(0,\tilde\omega_{0}^{2})}{\partial m_{0}^{}} $ at $\mu=0$ as functions of temperature $T$ (in \emph{solid line}, \emph{dashed line}, \emph{dotted line}, respectively).
} \label{fig:comp}
\end{figure}

As discussed in Section II, besides the $\frac{\partial \langle {\bar{q} q }\rangle }{\partial T}$, we have other definitions for the chiral susceptibility, such as $\frac{\partial \langle {\bar{q} q }\rangle }{\partial m_{0}^{}}$, even the simplified one in practical calculation, $\chi(0,\tilde\omega_0)=\frac{\partial B(0,\tilde\omega_0^2)}{\partial m_{0}^{}}$.
We should then check the equivalence of the critical temperature and the critical chemical potential determined with the different criteria.
In the region of first or/and second order phase transition, due to the functional relation among the dynamical mass, the quark condensate and the $B(0,\tilde\omega_0^2)$, the discontinuities of them are the same (the ones corresponding to the Wigner solution all emerge at the $T_{c,l} $ or/and $\mu_{c,l}^{}$,  and those relating to the Nambu solution all disappear at the $T_{c,h} $ or/and $\mu_{c,h}^{}$ ). All the chiral susceptibility defined in terms of $B(0,\tilde\omega_0^2)$ and $\langle \bar{q} q \rangle$ diverge at the respectively same $T_{c,l}^{}$ ($\mu_{c,l}^{}$), or $T_{c,h}^{}$ ($\mu_{c,h}^{}$). Therefore the critical states ($T_{c,l}^{}$, $\mu_{c,l}^{}$) ( ($T_{c,h}^{}$, $\mu_{c,h}^{}$) )  determined with different definitions of the chiral susceptibility are equivalent to each other for the first and second phase transitions.
While for crossover where the susceptibility does not diverge,  there are no rigorous manifestations  and just pseudo-critical condition (temperature, or/and chemical potential) to mark the chiral symmetry restoration. Different criteria might give different results, we should thus compare the results fixed with different definitions of the chiral susceptibility carefully.
To this end, we give the calculated temperature dependence of chiral susceptibility defined as $\frac{\partial \langle \bar{q} q \rangle }{\partial T}$, $\frac{\partial \langle \bar{q} q \rangle}{\partial m_{0}^{}}$ and  $\frac{\partial B(0, \tilde\omega_{0}^{2})}{\partial m_{0}^{}}$ at zero chemical potential in Fig.~\ref{fig:comp}.
The figure manifests evidently that the variation behaviors of the susceptibilities defined differently with respect to the temperature are generally almost the same, which exhibits an obvious peak, except for that  in terms of the quark condensate shows a little sharper than the other one. The continuous variation features of the susceptibility with all definitions confirm that the chiral phase transition is not a sharp phase transition but a crossover. The two susceptibilities defined as the derivative of  the current quark mass have their maxima at the same temperature demonstrates that these two criteria  are equivalent to each other, which give a pseudo-critical temperature $T_{c}^{}(\mu=0)=156.1\;$MeV.
While the susceptibility criterion defined as the derivative to temperature gives a pseudo-critical temperature  $T_{c}^{}(\mu=0)=150.8\;$MeV, which has been quoted as $T_{c}^{}$ in last paragraph.
These values are definitely highly consistent with the lattice QCD simulation results~\cite{Aoki:2006PLB,Bazavov:2012PRD,Ding:2014PRL}.
It is well known that the crossover means a smooth evolution from one phase to another, different criteria lead naturally distinct pseudo-critical temperature.
The about $5\;$MeV difference among the pseudo-critical temperatures determined with different definitions of the chiral susceptibility ({\it i.e.}, different criteria) is just a manifestation of the crossover.
Since the discrepancy among the pseudo-critical temperatures obtained with different definitions of the chiral susceptibility is quite small,
we will then implement the commonly taken definition of the chiral susceptibility (i.e., that defined as the derivative with respect to the temperature) as the signature to fix the phase diagram in the follows.

\begin{figure}[htb]
\centerline{\includegraphics[width=0.50\textwidth]{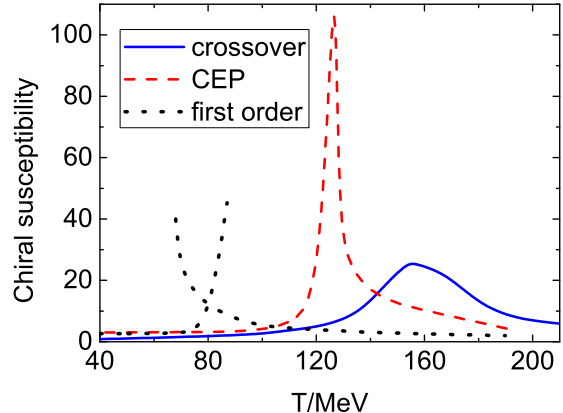}}
\caption{(color online) General characteristics of the chiral susceptibility in different regions
(\emph{solid}-- at $\mu_{q}^{} = 0 \;$MeV, {\it i.e.}, in crossover region; \emph{dotted}-- at $\mu_{q}^{} = 200 \;$MeV, {\it i.e.}, in first order transition region; \emph{dashed}-- at $\mu_{q}^{} = 85 \;$MeV, {\it i.e.}, near the chemical potential separating the crossover region from the first order transition region.)} \label{fig:chiral}
\end{figure}

Figs. \ref{fig:conden1} and \ref{fig:conden2} manifest apparently that the order parameter, scaled chiral quark condensate, behaves distinctly in the regions of different order phase transitions. In turn the chiral susceptibility demonstrates different features in the different regions.  The calculated temperature dependence of the chiral susceptibilities at three typical values of the chemical potential is shown in Fig.~\ref{fig:chiral}.
It is apparent that in the first order transition region, the susceptibility of the Nambu (DCSB) phase diverges at states different from those for that of the Wigner (CS) phase to diverge.
The region between the states for the susceptibilities of the two phases to diverge individually is
the coexistence region.
In the crossover region, the susceptibility is a smooth function involving a peak, {\it i.e.}, the susceptibility of the DCSB phase links with (in fact, changes to)  that of the CS phase not only continuously but also smoothly.
While, in the second order transition region, the susceptibility of the DCSB phase diverges
at the same location as that for the susceptibility of the CS phase to diverge.
%
%
The characteristic for the susceptibilities of the two phases to diverge at the same location defines thus the state which separates the crossover region from the first order transition region, namely, the CEP. Therefore the chiral susceptibility criterion can not only give the phase boundary of the chiral phase transition but also localize the position of the CEP.

\section{Deconfinement Phase Transition}

\subsection{Criterion for deconfinement transition}

The confinement is defined as that the color degrees of freedom are confined to the inside of hadrons and could not be observed as isolated states. It means that there does not exist an asymptotic free coloured state.
In turn, it can naturally be represented by the violation of the positivity of the spectral density function.
It has been shown that such a violation of the positivity associates the confinement with the dynamically-driven changes in the analytical structure of QCD's propagators and vertices~\cite{Gribov:1999EPJC,Munczek:1983PRD,Stingl:1984PRD,Cahill:1989AJP,Roberts:1992IJMPA,Hawes:1994rw,Pawlowski:2013PRD}.
To avoid the difficulty in calculating the spectral density function, one usually links it with the Schwinger function. The Schwinger function at finite temperature and finite chemical potential is defined as the Fourier transformation of the propagator~\cite{Bender:1996PRL,Bender:1998PLB,Mueller:2010EPJC,Qin:2011PRD,Gao:2014PRD,Qin:2013PRD,Bashir:2008PRC,Bashir:2009FBS}
 \begin{eqnarray}
&&D_\pm(\tau,|\vec{p}|=0)=T\sum_{n}e^{-i\omega_n\tau}S_\pm(i\omega_n+\mu,|\vec{p}|=0)     \quad  \notag\\
 & & \; = \int^{+\infty}_{-\infty}\frac{d\omega}{2\pi} {\rho_{\pm}^{}}(\omega,|\vec{p}|=0) \frac{e^{-(\omega+\mu)\tau}}{1+e^{-(\omega+\mu)/T}} \, ,
 \label{eq:Drho}
 \end{eqnarray}
where $S_{\pm}$ is the projected quark propagator defined as $S=S_{+} L_{+} + S_{-} L_{-}$ with  $L_{\pm} = \frac{1}{2}(1 \pm \gamma_{4})$, $\rho_{\pm}^{}$ is the corresponding spectral density function and is positive definite if the propagator contains an asymptotic state.
The violation of the positivity of the spectral density function is sufficient for determining the confinement~\cite{Roberts:1994PPNP,Alkofer:2001PR}, while the Schwinger function criterion  fails in some cases~\cite{Alkofer:2004PRD}. It is because the spectral density is integrated out in Schwinger
function, even though there's negativity in spectral density function, the Schwinger function  can still be positive after integrating.
The positivity of the Schwinger function is just a prerequisite but not a sufficient condition for
judging the deconfinement as found in the calculation in DS equation approach~\cite{Alkofer:2004PRD}.
It would then be much helpful if one can find a way for the Schwinger function, which is easy to calculate, to represent the properties of spectral density function directly.

Noticing that by differentiating the Schwinger function against the $\tau$, we have
\begin{eqnarray}
 &&D^{2n}_{\pm}(\tau,|\vec{p}|)=\notag \\
 && \quad \int^{+\infty}_{-\infty}\frac{d\omega}{2\pi}(\omega+\mu)^{2n} {\rho_{\pm}^{}}(\omega,|\vec{p}|)   \frac{e^{-(\omega+\mu)\tau}}
 {1+e^{-(\omega+\mu)/T}} \, , \qquad
 \label{eq:Dprime}
 \end{eqnarray}
where $D^{2n}$ is the $2n$ order derivative of $D$.
If discretizing the variables in Eq.~(\ref{eq:Dprime}) we have
$$D^{2n}(\tau_{l}^{})
=\sum_{m}^{} f^{n}(\omega_{m}^{},\tau_{l}^{})\rho(\omega_{m}^{}) \, , $$
then the $(n\times l)$-dimensional functions $D^{2n}(\tau_{l}^{})$ are able to determine the
$m = n\times l$ dimensional discretized spectral density function.
It can definitely reach to the continuum limit when $n \rightarrow \infty$.
This means that, even though the Schwinger function $D^0$ is not sufficient,
the series of $D^{2n}$ can determine the properties of the spectral density function completely.
If the spectral density function is positive definite, the series of $D^{2n}$ should be all positive,
and otherwise, when the spectral density function is somehow negative,
negativity will appear in the series of $D^{2n}$.

\subsection{Numerical Result}

\begin{figure}[htb]
\includegraphics[width=0.23\textwidth]{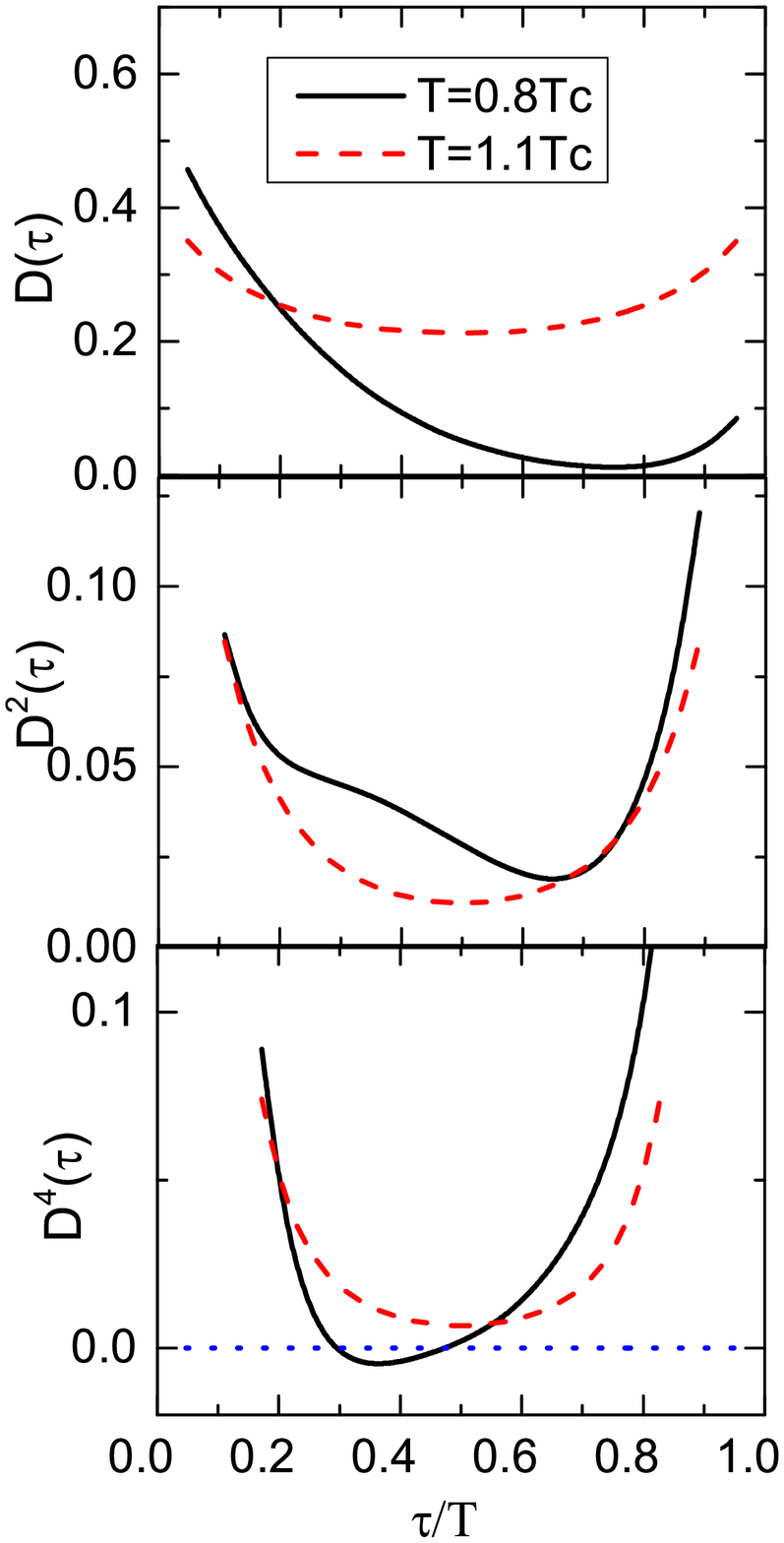}
\includegraphics[width=0.23\textwidth]{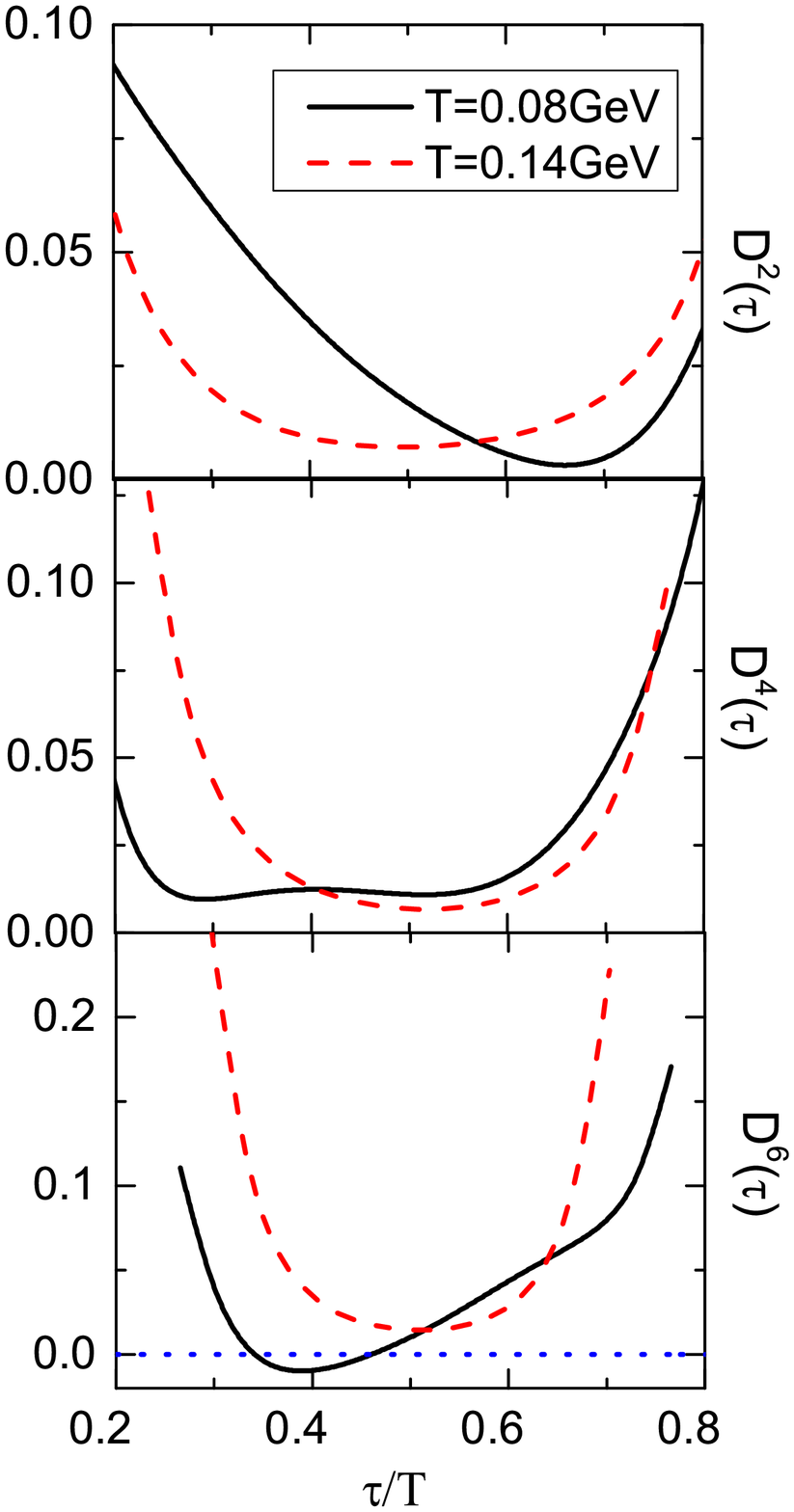}
\caption{ (color online) Calculated Schwinger function $D(\tau)$ and its derivatives $D^{2}(\tau)$, $D^{4}(\tau)$ at zero chemical potential
%
(\emph{left panel};  \emph{dashed}--$T= 1.1\, T_{c}^{\chi}$, \emph{solid}--$T= 0.8\, T_{c}^{\chi}$ )
and the $D^{2}(\tau)$, $D^{4}(\tau)$, $D^{6}(\tau)$ at chemical potential $\mu_{q}^{} = 110\;$MeV (\emph{right panel};
\emph{dashed}--$T= 140\;$MeV, \emph{solid}--$T= 80\;$MeV ). } \label{fig:decon}
\end{figure}

To show the validity of our criterion, we have calculated the $D(\tau)$, the second order derivative $D^2(\tau)$, the fourth order derivative $D^4(\tau)$ and the sixth order derivative $D^{6}(\tau)$ at many states $(T, \mu_{q})$.
The obtained results at $(1.1T_{c}, 0)$ and $(0.8T_{c}, 0)$ whose spectral density functions have been analyzed explicitly in Ref.~\cite{Qin:2013PRD}, and those at  $( 140, 110)\;$MeV and $(80, 110 )\;$MeV  are illustrated in Fig.~\ref{fig:decon}.

Fig.~\ref{fig:decon} manifests evidently that in case of zero chemical potential, the $D(\tau)$
at $T = 1.1\, T_{c}^{\chi}$ is positive definite, which is consistent with the positivity of the spectral density function (see, {\it e.g.}, Ref.~\cite{Qin:2013PRD}).
The $D(\tau)$ at $T=0.8\, T_{c}^{\chi}$ is  positive definite too, however the definite positivity of the spectral density function is violated (see, {\it e.g.}, Ref.~\cite{Qin:2013PRD}).
Inconsistence emerges in the Schwinger function criterion and the spectral density function criterion.
Nevertheless, the $D^{4}(\tau)$ accords with the spectral density function excellently.
In case of chemical potential $\mu_{q}^{} = 110\,$MeV ( $\mu_{B}^{} = 330\,$MeV), the $D(\tau)$, $D^{2}(\tau)$ even $D^{4}(\tau)$ are all positive, which could not illustrate the confinement nature
at $T=80\,$MeV. While positivity violation appears for $D^{6}(\tau)$.
It is then clear that analyzing the even order derivative of the Schwinger function (we refer it as the generalized Schwinger function hereafter) can play the role to identify the confinement--deconfinement phase transition efficiently.

From Fig.~\ref{fig:decon} one can also observe that the positivity violation of the generalized Schwinger function connects the change of the monotonicity of the function.
In general principle, if the $D(\tau)$ and its $2n$ order derivatives are all convex function,
they and the spectral function are positive definite and manifest the deconfinement.
While any concave behavior appears in $D^{2n}(\tau)$, positivity violation emerges for the $D^{2(n+1)}(\tau)$ and the spectral density function, which means a confinement.

\section{Phase Diagrams and Critical End Point}

\begin{figure}[htb]
\centerline{\includegraphics[width=0.50\textwidth]{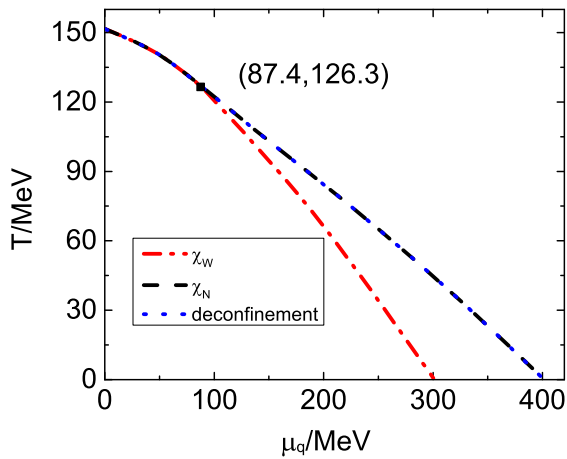}}
\caption{(color online) Calculated phase diagrams on the $T$--$\mu_{q}^{}$ plane
(\emph{dot--dashed}--Wigner chiral phase transition; \emph{dashed}--Nambu chiral phase transition;
\emph{dotted}--Deconfinement phase transition.)} \label{fig:phase}
\end{figure}

With the solutions of the quark's DS equation, we can take the chiral susceptibility criterion and the generalized Schwinger function criterion to give the complete phase diagrams in the $T$--$\mu$ plane.
For chiral susceptibility, we perform our calculations with definitions $\frac{\partial \langle \bar{q} q \rangle}{\partial T}$,  $\frac{\partial \langle \bar{q} q \rangle}{\partial m_{0}^{}}$ and  $\frac{\partial B(0, \tilde\omega_{0}^{2})}{\partial m_{0}^{}}$.
The calculated results with each of the definition show that in low chemical potential region,
the line demonstrating the states for the susceptibility of the Nambu phase to take its maximum overlaps with the line for that of the Wigner phase to take its maximum. However the two lines separate from each other in high chemical potential region.
This indicates that chiral phase transition in high chemical potential region is a first order phase transition but that in low chemical potential region is in fact a crossover,
just as mentioned in Section \uppercase\expandafter{\romannumeral3}.
And there exists a CEP to separate the two regions.
The obtained chiral phase diagram with the conventional definition of the chiral susceptibility $\frac{\partial \langle \bar{q} q \rangle}{\partial T}$ and that of the confinement phase diagram are displayed in Fig.~\ref{fig:phase}.
It is evident that the presently obtained phase diagram is qualitatively the same as the previous results.

%
Looking over the numerical data, we notice that our present calculation with the refined quark-gluon interaction vertex (CLR model) gives the position of the CEP at $(\mu_{E}^{B},T_{E}^{})=(262.3,\, 126.3)\;$MeV ($\mu_{E}^{B} = 3 \mu_{E}^{q}$), which yields the ratios ${T_{E}^{}}/{T_{c}^{}} = 0.84$, $\mu^{B}_{E}/T_{c}^{}=1.74$ and $\mu^{B}_{E}/T_{E}^{}=2.08$, which agree with the lattice QCD simulation results~\cite{Gavai:2005PRD,Schmidt:2008JPG,Li:2011PRD,Gupta:2014PRD}  very well.
Comparing such a result with our previous results in case beyond chiral limit~\cite{Gao:2016PRD,Xin:2014PRDb},
one can observe that both the chemical potentials and the temperatures of the CEP determined via different truncation schemes change accordingly (even though the amplitudes are different).
 To investigate the parameter dependence of the location of the CEP, we have performed a series calculations with maintaining the same quark condensate at $T= \mu = 0$. Our calculated results indicate that, to keep the quark condensate the same, the parameter $\omega$ in the interaction can take different values but the ${\mathcal{D}}$ should be almost a constant.
Moreover, as the parameter $\omega$ decreases, the $\mu_{E}^{}$ decreases and the $T_{E}^{}$ increases. Some of the concrete data are listed in Table~\ref{tab:Paradep-CEP}.
\begin{table}[htb]
\caption{Calculated parameter dependence of the location of the CEP $({\mu_{E}^{B}}, {T_{E}^{}})$ in case the quark condensate $\langle \bar{q}q \rangle_{\mu=\rm 1GeV}$ maintaining a constant ($(240\,\textrm{MeV})^{3}$) via different truncation schemes (All the dimensional quantities are in unit GeV) .}
\begin{tabular}{c|ccccc}
\hline \hline
Scheme & ~$(D\omega)^{1/3} $~ & $\omega$ & $T_{c} $ & $({\mu_{E}^{B}}, {T_{E}^{}})$ & $\mu_{E}^{B}/T_E$ \\
\hline
{} & $0.800$ & $0.450$ & $0.1562$ & $(0.2055, 0.1524)$ & $1.3572$ \\
RL  & $0.800$ & $0.500$ & $0.1503$ & $(0.3324, 0.1283) $ & $2.5908$ \\
~ & $0.800$ & $0.550$ & $0.1343$ & $(0.4371, 0.1143) $ & $3.8241$ \\
\hline
{} & $0.501$ & $0.450$ & $0.1536$ & $(0.1176, 0.1513) $ & $0.7773$ \\
Refined & $0.520$ & $0.500$ & $0.1508$ & $(0.2623, 0.1263) $ & $2.0768$ \\
~ & $0.536$ & $0.550$ & $0.1274$ & $(0.3909, 0.1022) $ & $3.8247$ \\
\hline \hline
\end{tabular}
\label{tab:Paradep-CEP}
\end{table}
Extending the idea discussed in Refs.~\cite{Qin:2011PRL,Xin:2014PRDa,Xin:2014PRDb},
one can infer that the $\frac{1}{\omega}$ plays the role of the radius of the interaction sphere.
An increase of the interaction radius ({\it i.e.}, the volume of the interaction sphere) plays the same role as increasing the density of the system,  it compensates then the effect of increasing the chemical potential. As a consequence, the $\mu_{E}^{B}$ decreases and the $T_{E}^{}$ increases simultaneously due to the compensation.

If we consider only the crossover region of the chiral phase transition, we can fit the chiral phase boundary line with an expansion formula:
\begin{equation}
T_c(z=\mu/T_{c,\mu})= T_{c,\mu=0}(1-\kappa \, z^{2} ),
\end{equation}
with $\kappa = 0.339$ if $\mu$ stands for that of the quarks,
and it is $\kappa_{B}^{}=0.038$ if $\mu$ refers to the baryon chemical potential.
This value is merely one time larger than that given in recent lattice QCD simulations, which
are typically around $0.018$ ~\cite{Endrodi:2011JHEP,Cea:2014PRD,Kaczmarek:2011PRD},
but consistent with other DSEs calculations~\cite{Fischer:2014NPA,Xin:2014PRDb,Gao:2016PRD}.
Since lattice QCD simulation suffers from the sign problem at finite chemical potential
and thus usually employs the techniques like Taylor expansion or analytic continuation,
we believe it would be reliable to determine the curvature parameter
through the DS equation approach which does not involve further approximation
when carrying out the calculation on finite chemical potential domain.

In the first order transition region, people usually take the phase equilibrium condition $P_{N}^{} = P_{W}^{}$ to fix the boundary line. Even though we could not give the boundary line in such a way now since we could not get the thermodynamical potential due to including the nonperturbative effect more generally in the presently refined truncation scheme, one can infer that the line must locate in the between of the (red) dot-dashed line and the (black) dashed line. From the condition  $P_{N}^{}=P_{W}^{}$  we directly have:
\begin{eqnarray}
\frac{d(P_{N}^{}-P_{W}^{})}{dT} & = & \big( \frac{\partial P_{N}^{}}{\partial \mu}-\frac{\partial P_{W}^{}}{\partial \mu} \big) \frac{\partial \mu}{\partial T} + \big( \frac{\partial P_{N}^{}}{\partial T} - \frac{\partial P_{W}^{}}{\partial T} \big)  \nonumber \\   & = & 0  \, .  \nonumber
\end{eqnarray}
With the thermodynamical relations $n=\partial P/\partial \mu$ and $s=\partial P/\partial T$, where $n$ is the particle number density and $s$ is the entropy density, we observe:

\begin{equation}
\frac{\partial \mu}{\partial T}=-\frac{s_{N}^{} - s_{W}^{}}{n_{N}^{} - n_{W}^{}}   \, .
\end{equation}
Typically, the CS phase has a larger entropy density and a larger quark number density,
which reads $n_{W}^{} > n_{N}^{}$ and $s_{W}^{} > s_{N}^{}$.
Therefore, there should be $\frac{\partial \mu}{\partial T} < 0$.
It indicates that the phase boundary line $\mu_{q}^{}(T)$ appears generally a monotonically decreasing line, and could not involve any backbend.

For the confinement--deconfinement phase transition, our presently calculated result, as shown in Fig.~\ref{fig:phase}, manifests that the dotted curve determined with $D^{4}(\tau) = 0$
or $D^{6}(\tau) = 0$ overlaps with the chiral phase transition line of the Nambu phase.
It indicates that the deconfinement phase transition coincides with the complete chiral symmetry restoration (chiral phase transition) exactly, and it makes the locations of the two kind CEPs coincide.
Meanwhile, the Wigner phase is always in deconfinement phase even when it is unstable
in low temperature and low chemical potential region.
This feature provides evidence again for that there does not exist hierarchy between
the chiral phase transition and the deconfinement phase transition.

In addition, if we employ the parametrization of the chemical freeze out condition reported in Ref.~\cite{Cleymans:2006PRC}, we would propose that the CEP may appear in the matter generated by the Au--Au collision
at the energy $\sqrt{s} \cong 14.6\;$GeV, or with the parametrization in Ref.~\cite{Andronic:2006NPA}, we obtain $\sqrt{s} \cong 13.9\;$GeV. If we take further the finite size effect into accounts~\cite{Chen:2015mga}, the energy to generate the CEP in the Au--Au collision will shift to $\sqrt{s} \cong 9.4\;$GeV, fitting the data from lattice QCD simulation~\cite{Borsanyi:2014PRL} with the energy dependence expression $\mu_B=d/(1+e\sqrt{s})$, we can also find the CEP is at the energy around $\sqrt{s} \cong 10\;$GeV.
Such an energy range $9 \sim 15\;$GeV is consistent with what the oscillation structure of the net baryon number fluctuation observed in recent RHIC experiments~\cite{Luo:2015ewa} hint.

\section{Summary }

With a refined truncation scheme of the Dyson-Schwinger equations developed recently, we studied the QCD phase transitions in this paper.
For the chiral phase transition, we analyzed the equivalence of the (generalized) chiral susceptibility criterion and the thermodynamical potential criterion.  The chiral susceptibility
criterion is much more powerful in case of taking the nonperturbative nature of the phase transitions into account where the thermodynamical potential is not available.
We also investigated the consistency of the chiral susceptibility criterion with distinct definitions
of the susceptibility ({\it i.e.}, that in different directions), and showed that
the susceptibility along different directions behaves in the same manner in the first order transition region but slightly differently in the crossover region.
For the deconfinement phase transition, we gave a generalized Schwinger function criterion, and  proved that the positivity violation of the generalized Schwinger function is definitely a sufficient condition to identify the confinement.

With these criteria and the solutions of the refined DS equations, we obtained the complete phase diagram of not only the chiral phase transition but also the deconfinement transition.
The results indicate that the two kind phase transitions coincide with each other completely.
Our results for the chiral phase transition predict that there exists a CEP and it locates at $(\mu_{E}^{B},T_{E})=(262.3,126.3)\;$MeV,  with $T_{E}^{}/T_{c}=0.84$ and $\mu^{B}_{E}/T_{c}=1.74$, which agrees with lattice QCD simulation results and former DS equation results very well.
The obtained phase boundary coincides not only with the lattice QCD results and previous DS equation results well, but also the feature in general principle.
With the parametrization for the collision energy dependence of the chemical potential we propose that the CEP may appear in the states generated by the $\sqrt{s} \cong 9 \sim 15\;$GeV Au-Au collision.

Comparing our presently obtained results with the refined truncation scheme and the (previous) ones with the bare vertex approximation, one can notice that there does not exist obvious discrepancy between the results via different truncation schemes in general. However the refined scheme shifts the location of the CEP to lower chemical potential and higher temperature. It provides evidence for that the CEP locates in the region of the states being able to generated in the presently planned experiments. Nevertheless massive works are required to detect the CEP since it locates at the phase boundary line but what experiments can observe are those after the chemical freeze out.
Exploring the chemical freeze out conditions with the refined truncation scheme is thus under progress.

\bigskip


The work was supported by the National Natural Science Foundation of China under Contracts No. 11435001; the National Key Basic Research Program of China under Contract No. G2013CB834400 and
2015CB856900.

\end{document}